\documentclass[twocolumn,twoside,slac_two]{revtex4}
\usepackage{graphicx}
\usepackage{fancyhdr}
\begin{document}

\title{Hot Results from CLEO-c}

%

\author{John M. Yelton}
\affiliation{Physics Department, University of Florida, Gainesville, FL 32611}

\begin{abstract}
I briefly review some of the latest results obtained using the CLEO-c detector.
\end{abstract}

\maketitle

\thispagestyle{fancy}


\section{Introduction}
The CLEO detector has been taking data at the Cornell Electron 
Storage Ring since 1980. 
Its latest (and last) configuration is CLEO-c, which is optimized 
for running in the charm
threshold region. The main differences between it and the CLEO III 
configuration, is the 
replacement of the silicon vertex detector with a lightweight inner
 drift chamber. The 
magnetic field is now 1T. These two changes both help with the 
measurement of low momentum tracks.

The main datasets taken by CLEO-c are a) 
around 54$\ pb^{-1}$ taken at the $\psi(2S)$ 
resonance (corresponding to around 27,000,000 $\psi(2S)$ decays), 
b) a large block 
of running at the $\psi(3770)$, of which 281$\ pb^{-1}$ is processed, 
more is already
taken, 
and the target is 670$pb^{-1}$, c) running at just above 
4170 MeV, which is just above the $D_sD_s^*$,
of which most analyses use $\approx195\ pb^{-1}$, some use
$\approx 300\ pb^{-1}$, and more data
will be taken (with a target of 720$pb^{-1}$), 
and d) an energy scan in the region
3970-4260 MeV. 
Note, that there is no longer any running planned at the $J/\psi$ energy,
although some $J/\psi$ studies can be made using the $\psi(2S)$ decays. 
The data-taking 
program will end in Spring 2008.

\section{$D^0$ and $D^+$ Hadronic Branching Fractions}

There have been many measurements of the D hadronic branching fractions, but 
to there is no environment can compete with an $e^+e^-$ collider running at
the $\psi(3770)$ energy. CLEO has published these fractions 
using $56\ pb^{-1}$ of data \cite{BF1}; 
here I show preliminary results using $281\ pb^{-1}$ of data. 
This second analysis has been presented in detail elsewhere \cite{BF2}. 
We already have doubled this dataset, but that new 
data has not yet been analyzed, and yet more is on the way. 

The basic technique depends on the fact that if there is one $D$ meson
in the event, there must be one $\bar{D}$ meson, and nothing else. By 
first cutting on $\Delta(E)=E_{D-TAG}-E_{BEAM}$, we can plot
$M_{BC}=\sqrt{E^2_{BEAM}-p^2_{D-TAG}}$ and the signals are spectacularly
clean. We then calculate the ratio of doubly-tagged events 
(those where both the $D$ and $\bar{D}$ were constructed), 
to singly-tagged events (those where only one of the two is 
reconstructed), we can extract the 
absolute branching fractions for 3 $D^+$ decays and 6 $D^0$ decays.

The results of this analysis are shown in Table~I. Further comparisons
with the rest of the world data are shown elsewhere \cite{BF2}. 
Note that these results will supercede those of the previous analysis.

\begin{table}[h]
\begin{center}
\caption{$D$ Hadronic Branching Fractions}
\begin{tabular}{|c|c|c|}
\hline 
{\textbf{Particle}} & {\textbf{Decay}} & {\textbf{Branching Fraction (\%)}} \\
\hline $D^0$ & $K^-\pi^+$ & $3.88\pm0.04\pm0.09$ \\
\hline $D^0$ &$K^-\pi^+\pi^0$ & $14.6\pm0.1\pm0.4$ \\
\hline $D^0$ &$K^-\pi^+\pi^-\pi^+$ &$8.3\pm0.1\pm0.3$ \\
\hline
\hline $D^+$ &$K^-\pi^+\pi^+$ &$9.2\pm0.1\pm0.3$ \\
\hline $D^+$ &$K^-\pi^+\pi^+\pi^0$ &$6.0\pm0.1\pm0.2$ \\
\hline $D^+$ &$K^0_s\pi^+$ &$1.55\pm0.02\pm0.05$ \\
\hline $D^+$ &$K^0_s\pi^+\pi^0$ &$7.2\pm0.1\pm0.3$ \\
\hline $D^+$ &$K^0_s\pi^+\pi^-\pi^+$ &$3.13\pm0.05\pm0.14$ \\
\hline $D^+$ &$K^-K^+\pi^-$ &$0.93\pm0.02\pm0.03$ \\
\hline
\hline
\end{tabular}
\label{BF1}
\end{center}
\end{table}

I would just like to mention that CLEO-c 
has also recently made the most 
precise measurement of the $D^0$ mass \cite{D0mass}. 
This was made using the decay mode
$D^0 \to K^0_s \phi$. The result is 
$M(D^0)=1864.847\pm0.150\pm0.095$ MeV.
The result is particularly important because it leads to the
conclusion that 
the binding energy of the X(3872) when interpreted as a $D\bar{D}$ molecule
is $0.6\pm0.6$ MeV.

\section{$D_s$ Branching Fractions}

It is fitting that CLEO is still interested in the $D_s$, as it was 
responsible for the discovery of the particle in 1984 \cite{DSD}. 
That, and many other measurements in this sector, 
were performed using $e^+e^-$ collision
energies in the $\Upsilon$ region. 
Now, we have the chance to work at
a little above $D_s$ threshold. 
First, a scan was taken in the energy
range 3.97-4.26 GeV. 
It was found that the optimal place to operate 
is 4.17 GeV, and so $314\ pb^{-1}$ of data has been taken there.
At each energy of the scan, 
the cross section for $D_sD_s,D_s^*D_s$,
and $D_s^*D_s^*$ was calculated along with the cross section of $D\bar{D}$.
These cross-sections are interesting in themselves,
and details can be seen elsewhere \cite{RP}.

At the energy of 4.17 GeV, the majority of the $D_s$ mesons are produced
via $D_sD_s^*$. This produces a complication not present in the non-strange
case, as there is a low-energy photon in the event 
as well as the two mesons we are 
interested in. However, there is good kinematic separation between the modes.
This analysis (based on $195\ pb^{-1}$ of data), is presented in more
detail elsewhere \cite{DS}. Here I just present the $D_s$ absolute
branching fractions in Table~II.

\begin{table}[h]
\begin{center}
\caption{$D_s$ Hadronic Branching Fractions}
\begin{tabular}{|c|c|c|}
\hline 
{\textbf{Particle}} & {\textbf{Decay}} & {\textbf{Branching Fraction (\%)}} \\
\hline $D_s^+$ & $K_s^0K^+$ & $1.50\pm0.09\pm0.05$ \\
\hline $D_s^+$ &$K^-K^+\pi^+$ & $5.57\pm0.30\pm0.19$ \\
\hline $D_s^+$ &$K^-K^+\pi^+\pi^0$ &$5.62\pm0.33\pm0.51$ \\
\hline $D_s^+$ &$\pi^-\pi^+\pi^+$ &$1.12\pm0.08\pm0.05$ \\
\hline $D_s^+$ &$\pi^+\eta$ &$1.47\pm0.12\pm0.14$ \\
\hline $D_s+$ &$\pi^+\eta^{\prime}$ &$4.02\pm0.27\pm0.30$ \\
\hline
\hline
\end{tabular}
\label{BF2}
\end{center}
\end{table}

Note that we do not include the traditional normalizing mode for 
$D_s$ decays, namely $\phi\pi^+$. This is because there is a non-trivial
background to the $\phi\to K^+K^-$ signal, 
and different experiments have chosen different
$\phi$ mass cuts, leading to different amounts of this background being
included. 
The complicated substructure of this decay mode make it a candidate for an
amplitude analysis rather than using it as a normalizing mode.

\section{Decay modes of $D_s$ into two pseudo-scalars}

The analysis technique here is straightforward as we look for single
$D_s$ decays in the 4170 MeV data (note that in this $preliminary $ analysis,
$\approx 300 pb^{-1}$ are used). We search for four Cabibbo 
suppressed modes ($\pi^0K^+,K^+\eta,K^+\eta^{\prime}$), one
decay that is expected to be forbidden ($\pi^+\pi^0$), and compare these
with three Caibibbo-allowed decays measured in the same dataset ($\pi^+\eta$,
$\pi^+\eta^{\prime}$, and $K^+K^0$).
The $preliminary$ results for the branching 
ratios, seen in public for the first time are:

$(D_s\to K^+\eta)/(D_s\to\pi^+\eta)=0.080\pm0.015$

$(D_s\to K^+\eta^{\prime})/(D_s\to\pi^+\eta^{\prime})=0.039\pm0.013$

$(D_s\to K^0\pi^+)/(D_s\to K^+K^0)=0.083\pm0.009$

$(D_s\to K^+\pi^0)/(D_s\to K^+K^0)=0.042\pm0.012$

$(D_s\to \pi^+\pi^0)/(D_s\to K^+K^0) < 0.04$

These results are $preliminary$ 
statistics dominated, and more statistics will be available.

\section{Charmed meson decay constants}

Measurement of the decay constants $f_{D^+}$ and $f_{D_s}$ are of great 
interest, but are very difficult experimentally because they require 
investigation of modes with neutrinos in their final state. CLEO-c
has published analyses of $D^+ \to \mu^+\nu$ and $D^+ \to e^+\nu$ \cite{2fD},
and also $D^+ \to \tau^+\nu$ \cite{newfD}, each with the $281pb^{-1}$ dataset. 
The analysis of $D_s^+\to\mu^+\nu$ and $D_s^+ \to \tau^+\nu, \tau \to \pi^+\nu$
has now been sent for publication
using $314pb^{-1}$ \cite{newDs}, 
and details can be found there. The second $D_s$ analysis,
using the decays of the $\tau$ into electrons, 
is complementary to the first and 
largely independent of it.
The value for $f_D$ is found to be $(223\pm 17 \pm 3)\ $ MeV, the combined
value for $f_{D_s}$ is $(273\pm10\pm5)\ $MeV, and the ratio,
$(f_{D_s}/f_D)=1.22\pm0.09\pm0.03$ (these results are all preliminary).
This ratio is very consistent with most 
models, including recent lattice QCD models.

\section{$\psi(2S)$ data}

There are already some results shown using $\psi(2S)$ data, demonstrating
its use as a factory for $\chi_c$ production. 
We now have an order of magnitude more data taken. This
will enable us to study the decays of all three $\chi_c$ states, 
the $\eta_c$ and
the $h_c$ in unprecedented detail. It is also possible to do some
detailed analysis of $J/\psi$ decays found from the di-pion decays of
the $\psi(2S)$.
Just one example of $\chi_c$ physics is the 
decays into $KK\pi\pi$, which 
can be seen in six different charge combinations, 
each with
good signal to noise ratios. It will take a lot of work to understand the rich
resonant substructure of these decay modes. These are just a few of the 50 or 
so $\chi_c$ decay modes that we can measure,
and we are limited only by the manpower
to do the necessary analyses.

\section{Conclusions}

I have presented a series of results, most preliminary and some final,
from the data taken by the CLEO-c detector configuration. With much more data
becoming available, please expect many more results in the 
next months and years.  

\begin{acknowledgments}
I thank my collaborators in CLEO for their help in preparing this talk, 
in particular
Hanna Mahlke, Istvan Danko and Mikhail Dubrovin.
\end{acknowledgments}

\bigskip 

\end{document}